\documentstyle[12pt,epsfig]{article}

\newcommand{\Red}{}
\newcommand{\Blue}{}

\newcommand{\Pslash}{\kern 0.2 em P\kern -0.56em \raisebox{0.3ex}{/}}
\newcommand{\pslash}{\kern 0.2 em p\kern -0.4em /}
\newcommand{\Sslash}{\kern 0.2 em S\kern -0.56em \raisebox{0.3ex}{/}}
\newcommand{\ul}{\underline}
\newcommand{\tstrut}{\rule[-2.4mm]{0mm}{8mm}}
\newcommand{\nn}{\nonumber}
\newcommand{\itemsym}{\Red{\Large{$\bigcirc\hspace{-5.8mm}\triangle$}}\quad}
\newcommand{\negspace}{\vspace{-1.35mm}}
\newcommand{\bm}[1]{\mbox{\boldmath $#1$}}

\begin{document}

\vspace*{-8mm}
{\small
hep-ph/9710496 \hfill VUTH 97-18\\
NIKHEF 97-043 \hfill FNT/T-97/13}\\
\begin{center}
{\LARGE\bf Estimates~for~non-leading~distribution~functions\footnote{Talk 
presented at the workshop 'Deep Inelastic Scattering off Polarized 
Targets: Theory Meets Experiment', DESY Zeuthen, Germany, Sept. 1-5, 1997}}

%\vspace{1cm}
\vspace{8mm}
{\underline{R.~Jakob}$^{a,b}$, P.J.~Mulders$^{b,c}$, J.~Rodrigues$^{b,d}$}

%\vspace*{1cm}
\vspace*{6mm}
{\it $^a$
%Dipartimento di Fisica Nucleare e Teorica, 
Universit\`{a} di Pavia and INFN, Sezione di Pavia, 
Via Bassi 6, 27100 Pavia, Italy}\\ 

%\vspace*{3mm}
\vspace*{1mm}
{\it $^b$NIKHEF, P.O.Box 41882, NL-1009 Amsterdam, the Netherlands}\\

%\vspace*{3mm}
\vspace*{1mm}
{\it $^c$
%Department of Physics and Astronomy, 
Free University, De Boelelaan 1081, NL-1081 HV Amsterdam, the Netherlands}\\

%\vspace*{3mm}
\vspace*{1mm}
{\it $^d$Instituto Superior T\`{e}cnico, Av.~Rovisco Pais, 100 Lisboa Codex,
Portugal }\\

%\vspace*{2cm}
\vspace*{6mm}

\end{center}

%%%%%%%%%%%%%%%%%%%%%%%%%%%%%%%%%%%%%%%%%%%%%%%%%%%%%%%%%%%%%%%%%%%%%%%%
\begin{abstract}
\noindent
Estimates for leading and non-leading `twist' distribution functions are
obtained within the framework of a diquark spectator model using a 
non-local operator representation.  
\end{abstract}

%%%%%%%%%%%%%%%%%%%%%%%%%%%%%%%%%%%%%%%%%%%%%%%%%%%%%%%%%%%%%%%%%%%%%%%%
More details about the method and the results reported here can be found in
the long write-up~\cite{jmr97}.\\[-8mm]

\section{introduction}

\vspace{1mm}
\noindent
%% ---------   Slide 2 ------- quark-quark correlation functions

In a field-theoretic description of hard scattering processes the
information on the hadronic structure is contained in matrix elements
of non-local operators (parton correlation functions). For instance, the
distribution functions of quarks in hadrons are obtained from these hadronic 
matrix elements by tracing them with certain Dirac matrices and integration
over components of the quark momentum.\\ 
Distribution functions are universal in the sense, that they occur in the same
form in the description of all hard processes involving the same kind of soft
physics. Being, however, of genuine non-perturbative nature, in general, they
can not be calculated by perturbative means (with the exception of heavy quark
distribution functions). Ultimately, one could wait for lattice gauge theory to
provide the answer, but regarding the enormous difficulties those
techniques encounter, in the meanwhile, model estimates for the functions may
be very useful in predicting cross sections or asymmetries in future
experiments. 

\section{correlations and distribution functions}

\itemsym \ul{quark-quark correlation function}\\[2mm]
The simplest correlation function containing information on the distribution
of a quark (with momentum $p$) inside a hadron (characterized by momentum $P$
and spin vector $S$) can be written for each flavor 
as~\cite{sop77,cs82,jaf83}\\[-9mm]
\begin{center}
\includegraphics[width=4.5cm]{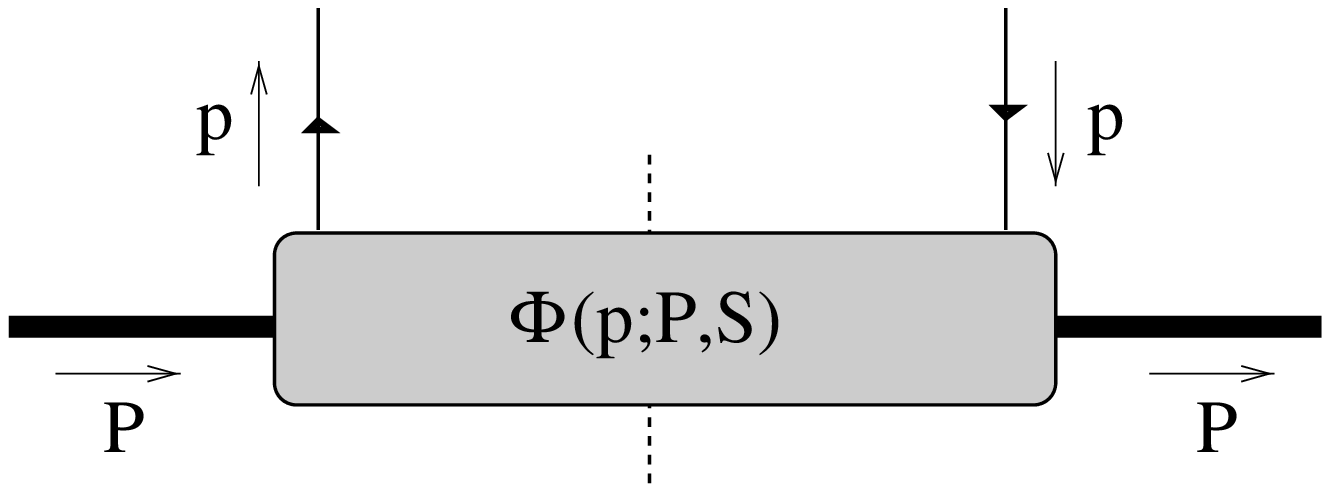}
\end{center}
\begin{equation}
\label{eq:Phi}
\Phi_{ij}(p,P,S)=\sum_X\int\frac{d^4x}{(2\pi)^4}\;
e^{ip\cdot x}\;\langle P,S|\Red{\overline\psi_j(0)}|X\rangle
\langle X|\Red{\psi_i(x)}|P,S\rangle.
\end{equation} 

%% ---------   Slide 3 -------  inclusive lepton hadron

\itemsym \ul{inclusive lepton-hadron}\\[2mm]
We encounter the quark-quark correlation function in the field theoretical
description of, e.g. deep inelastic lepton hadron scattering, whose amplitude
squared is depicted by the famous hand-bag diagram (the usual invariants
indicated besides it)
\begin{center} 
\includegraphics[width=5cm]{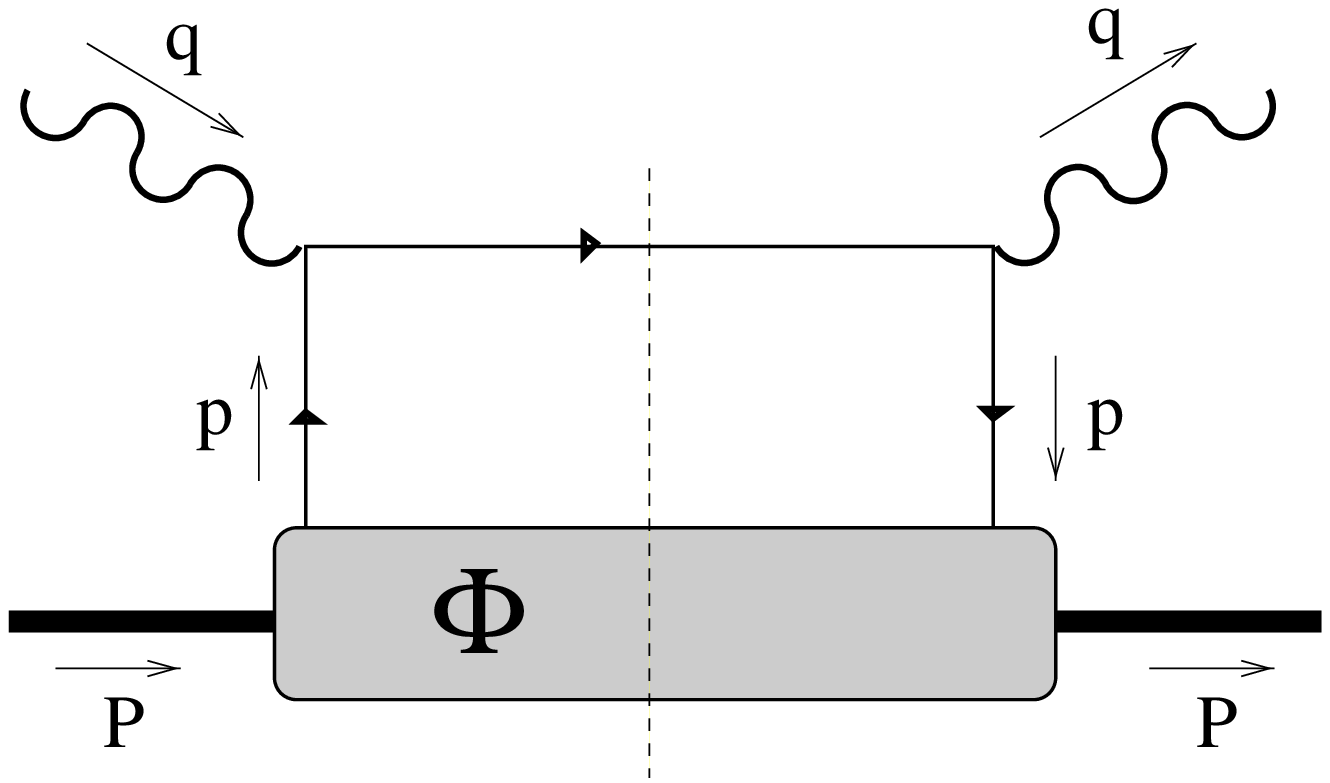}
\end{center}
\vspace{-34mm}
\begin{flushright}
\begin{minipage}{30mm}
\[ Q^2=-q^2 \]
\[ x_B={Q^2\over 2P\cdot q}. \]
\end{minipage}
\end{flushright}
\vspace{4mm}
In fact, in the cross section the correlation function occurs traced with a
Dirac matrix, $\gamma^+$, and integrated over 
three components of the quark momentum
\begin{equation} 
\Red{f_1(x)} = 
\left.\frac{1}{2}\int dp^-\,d^2\bm{p}_T\;{\rm Tr}(\Phi\gamma^+) 
\right|_{p^+ = x P^+}
\end{equation} 
which defines a {\em distribution function}, $f_1(x)$, depending only on the
longitudinal momentum fraction $x=p^+/P^+$. The parton model predictions for
the structure functions $F_1$ and $F_2$ are given (to lowest order) in terms 
of the distribution functions as 
\begin{equation} 
2F_1(x_B)={F_2(x_B)\over x_B}=\sum_a e_a^2\,\Red{f_1^a(x_B)}.
\end{equation} 

%% ---------   Slide 4 ------- Dirac projections

We introduce the notation
\begin{equation} 
\label{eq:disdef}
\Phi^{\Red{[\Gamma]}}(x) =
\left.{1\over 2}\int dp^-d^2\bm{p}_T\;
{\rm Tr}\left(\Phi\Gamma\right)\right|_{p^+=xP^+}
\end{equation} 
for the quark-quark correlation function traced with a certain Dirac matrix 
and integrated over $dp^-d^2\bm{p}_T$.\\

\itemsym \ul{leading twist distribution functions}\\[2mm]
Depending on the Dirac matrix involved, different spin properties of the
hadronic structure are probed. To leading order (in an $1/Q$ expansion) the 
following projections occur in the description of hard processes
($\lambda$ is the helicity of the hadron, $S_T$ the transverse part of the
spin vector) 
\begin{center} \fbox{\begin{minipage}{0.98\textwidth}
\begin{eqnarray}
\Phi^{[\Red{\gamma^+}]}(x)                    &=& \Red{f_1(x)}\\
&&\nn\\
\Phi^{[\Red{\gamma^+\gamma_5}]}(x)            &=& \lambda \; \Red{g_1(x)}\\
&&\nn\\
\Phi^{[\Red{i\sigma^{\alpha +}\gamma_5}]} (x) &=& S_T^\alpha \; \Red{h_1(x)}
\end{eqnarray}
\rule{0mm}{0mm}
\end{minipage}} \end{center} 
which have an intuitive probabilistic interpretation.\footnote{The notation of
the functions follows Refs.~\cite{jj92,tm96}.}
\newpage

\itemsym \ul{interpretation}\\[2mm]
\begin{itemize}
\item
$f_1(x)$ gives the probability of finding a quark with light-cone 
momentum fraction $x$ in the ``$+$''-direction (and any transverse momentum).
\begin{center} 
\includegraphics[width=2cm]{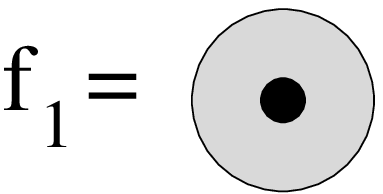}\\[4mm]
\end{center}
\item
$g_1(x)$ is a chirality distribution: in a hadron that is in a positive
helicity eigenstate, it measures the probability of finding a 
right-handed quark with light-cone momentum fraction $x$ minus the
the probability of finding a left-handed quark with the same light-cone
momentum fraction. 
\begin{center}
\includegraphics[width=5cm]{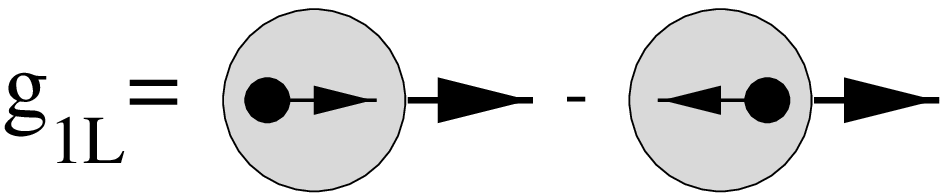}\\[1mm]
\end{center} 
\item
$h_1(x)$ is a transverse
spin distribution: in a transversely polarized hadron, it measures the
probability of finding quarks
with light-cone momentum fraction $x$ polarized along the direction of
the polarization of the hadron minus the probability of finding quarks
oppositely polarized. 
\end{itemize} 
\begin{center} 
\includegraphics[width=3.6cm]{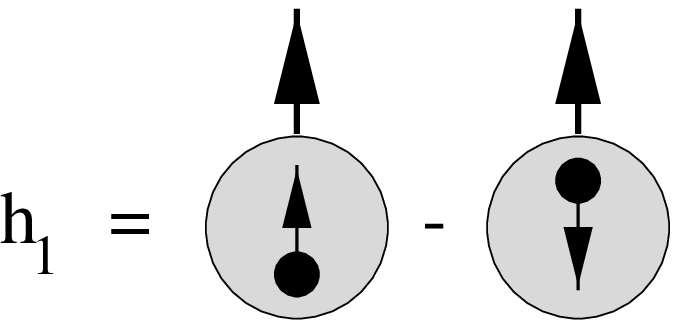}
\end{center} 

%% ---------   Slide 5 ------- higher twist functions

\itemsym \ul{twist 3 functions}\\[2mm]
The subleading (`higher twist') functions have no intuitive probabilistic
interpretation. Nevertheless, they are well defined via the quark-quark
correlation function traced with appropriate Dirac matrices. The pre-factor
$M/P^+$ behaving like $1/Q$ in a hard process signals the sub-leading
(i.e. `twist' 3) nature of the corresponding distribution functions
\begin{center} \fbox{\begin{minipage}{0.98\textwidth}
\begin{eqnarray} 
\Phi^{[\Red{1}]}(x)                     &=& {M\over P^+}\;\Red{e(x)}\\
\Phi^{[\Red{\gamma^\alpha\gamma_5}]}(x) &=& {M\over P^+}\;
                                            S_T^\alpha\;\Red{g_T(x)}\\
\Phi^{[\Red{i\sigma^{+-}\gamma_5}]} (x) &=& {M\over P^+}\;
                                            \lambda\;\Red{h_L(x)}
\end{eqnarray} 
\rule{0mm}{2mm} 
\end{minipage}} \end{center} 

%% ---------   Slide 6 -------  semi-inclusive lepton hadron

\itemsym \ul{transverse momentum}\\[2mm]
In some processes additional information is gained by considering observables
which depend on the transverse momentum of an external hadron (like
differential cross sections). Typically, these are processses with at 
least three external momenta, since those cannot be all collinear, in
general. The transverse momentum dependence of the observables is related 
to the transverse degrees of freedom of partons in the hadrons. Let us
examplify the relation for one hadron-inclusive lepton-hadron scattering (see
e.g.~\cite{tm96}). The amplitude (squared) is depicted by
\begin{center} 
\includegraphics[width=5cm]{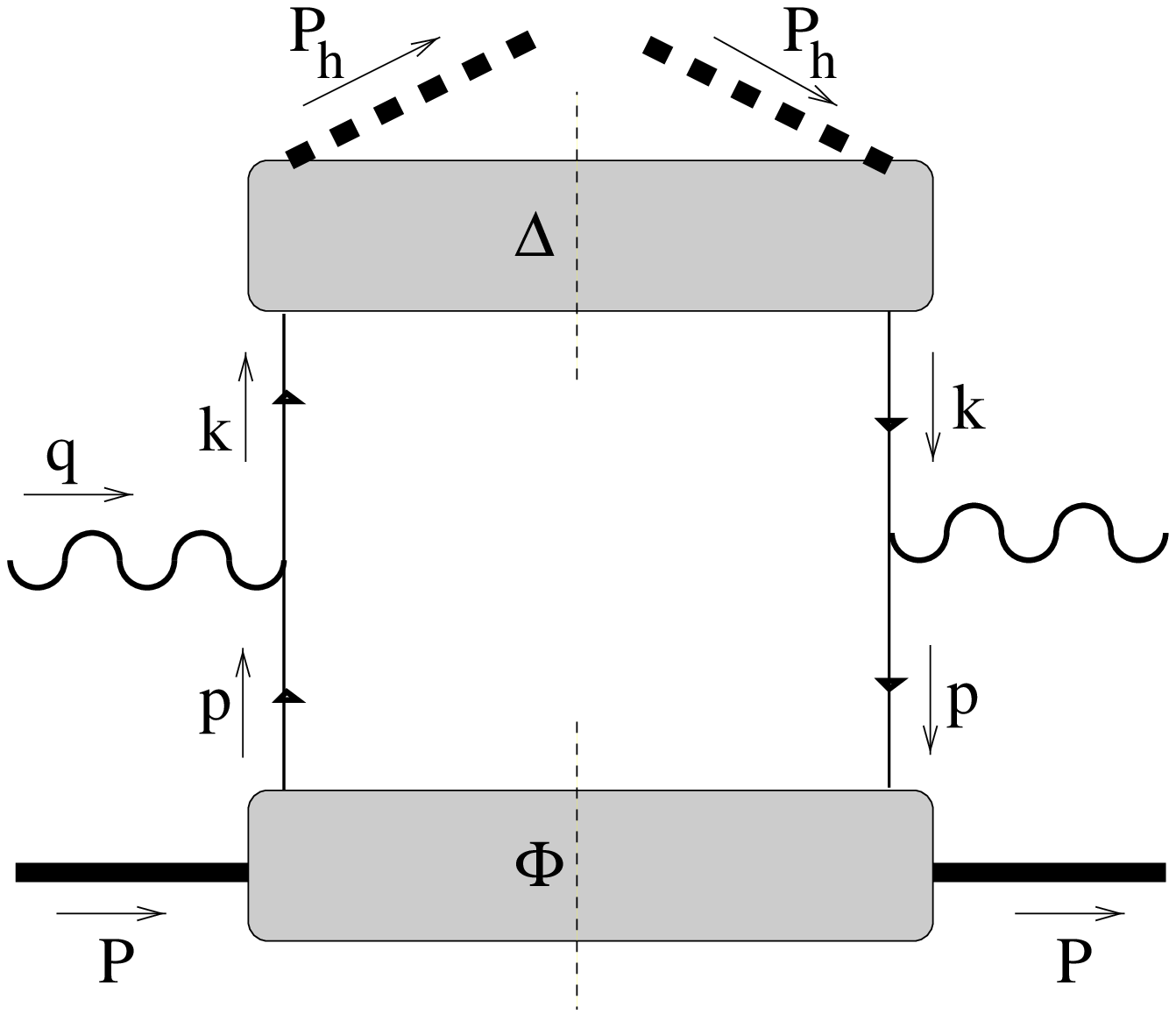}
\end{center}
\vspace{-43mm}
\begin{flushleft}
\begin{minipage}{40mm}
\hspace*{2mm}\large $[P_h^-,P_h^+,\bm{0}_T]$\\[6.6mm]
\hspace*{2mm}\large $[q^-,q^+,\bm{q}_T\neq 0]$\\[6.6mm]
\hspace*{2mm}\large $[P^-,P^+,\bm{0}_T]$
\end{minipage}
\end{flushleft}
\vspace{6mm}
In a frame where the target momentum, $P$, and the momentum of the observed
hadron in the outgoing channel, $P_h$, are collinear, the photon momentum will
have a 
transverse component (as indicated besides the diagramm).\footnote{The 
reasoning in other frames is similar, one of the external momenta unavoidably
will have a transverse component.}\\
The external transverse momentum, $\bm{q}_T$ is related via 
momentum conservation, i.e., a
$\delta(\bm{q}_T+\bm{p}_T-\bm{k}_T)$ function at the vertex, to the
transverse momentum 
components on the quark lines. Thus, observables differential in the 
transverse momentum will involve functions like
\begin{equation} 
\Red{f_1(x,\Blue{\bm{p}_T})} = 
\left.\frac{1}{2}\int dp^-\;{\rm Tr}(\Phi\gamma^+) 
\right|_{p^+ = x P^+, \Blue{\bm{p}_T}}
\end{equation} 
depending on the longitudinal momentum fraction {\em and} the transverse
momentum components which are not integrated out. E.g., the differential cross
section reveals a behavior
\begin{equation} 
\frac{d\sigma}{dx_B\,dy\,dz\,d^2\Blue{\bm{q}_T}}\sim
\int d^2\bm{p}_T \, d^2\bm{k}_T\,
\delta(\Blue{\bm{p}_T}+\Blue{\bm{k}_T}-\Blue{\bm{q}_T})\;
\Red{f_1(x_B,\Blue{\bm{p}_T})\;D_1(z,\Blue{\bm{k}_T})}
\end{equation} 
where $D_1(z,k_T)$ is a transverse momentum dependent fragmentation function
which describes the hadronization of the quark. In straightforward analogy we
define the functions
\begin{equation} 
\Phi^{\Red{[\Gamma]}}(x,\Blue{\bm{p}_T}) =
\left.{1\over 2}\int dp^- {\rm Tr}\left(\Phi\Gamma\right)
\right|_{p^+=xP^+,\Blue{\bm{p}_T}}
\end{equation} 
which differs from definition (\ref{eq:disdef}) only by the left-out
integration over transverse momentum.\\

\itemsym \ul{leading twist distribution functions} 
($\bm{p}_T$ dependent)\\[2mm]
Not surprisingly, there are more functions involved due to the additional
degrees of freedom. To leading `twist' there are six functions defined by
\begin{center} \fbox{\begin{minipage}{0.98\textwidth}
\begin{eqnarray}
\Phi^{[\Red{\gamma^+}]}(x,\Blue{\bm{p}_T})            &=& 
\Red{f_1(x,\Blue{\bm{p}_T^2})}\\
&&\nn\\
\Phi^{[\Red{\gamma^+\gamma_5}]}(x,\Blue{\bm{p}_T})    &=& 
\lambda \; \Red{g_{1L}(x,\Blue{\bm{p}_T^2})}
+\Red{g_{1T}(x,\Blue{\bm{p}_T^2})}\frac{\bm{p}_T\!\cdot\!\bm{S}_T}{M}\\
&&\nn\\
\Phi^{[\Red{i\sigma^{\alpha +}}]} (x,\Blue{\bm{p}_T}) &=& 
S_T^\alpha\;\Red{h_{1T}(x,\Blue{\bm{p}_T^2})}
+\frac{p_T^\alpha}{M}\left[\lambda\;
\Red{h_{1L}^\perp(x,\Blue{\bm{p}_T^2})}
+\Red{h_{1T}^\perp(x,\Blue{\bm{p}_T^2})}
\frac{\bm{p}_T\!\cdot\!\bm{S}_T}{M}\right]
\end{eqnarray}
\rule{0mm}{0mm}
\end{minipage}} \end{center} 
Of course, all those Dirac projections reduce after integration over
transverse momenta to the forms shown before, like for instance, 
$\int d^2\bm{p}_T \; \Phi^{[\gamma^+]}(x,\bm{p}_T)=\Phi^{[\gamma^+]}(x)$,
etc.\\[4mm] 

% ---------   Slide 7b  -------

\itemsym \ul{twist 3 functions} ($\bm{p}_T$ dependent)\\[2mm]
Similarly, there is a larger number of `twist 3' functions depending on $x$
and $\bm{p}_T$ which are obtained by the projections:
\begin{center} \fbox{\begin{minipage}{0.98\textwidth}
\begin{eqnarray}
\Phi^{[\Red{1}]}(x,\Blue{\bm{p}_T}) &=& 
\frac{M}{P^+}\,\Red{e(x,\Blue{\bm{p}_T^2})} \\
&&\nn\\
\Phi^{[\Red{\gamma^i}]}(x,\Blue{\bm{p}_T}) &=&
\frac{p_T^i}{P^+}\,\Red{f^\perp(x,\Blue{\bm{p}_T^2})}\\
&&\nn\\
\Phi^{[\Red{\gamma^i\gamma_5}]}(x,\Blue{\bm{p}_T}) &=&
\frac{M\,S_T^i}{P^+} \,\Red{g^\prime_T(x,\Blue{\bm{p}_T^2})}\nn\\&&
{}+\frac{p_T^i}{P^+}\left(\lambda \Red{g_L^\perp(x,\Blue{\bm{p}_T^2})}
+\frac{\bm{p}_T\!\cdot\! \bm{S}_T}{M}\Red{g_T^\perp(x,\Blue{\bm{p}_T^2})}
\right)\\
&&\nn\\
\Phi^{[\Red{i\sigma^{ij}\gamma_5}]}(x,\Blue{\bm{p}_T}) &=&
\frac{S_T^ip_T^j-S_T^jp_T^i}{P^+}\,\Red{h_T^\perp(x,\Blue{\bm{p}_T^2)}}\\
&&\nn\\
\Phi^{[\Red{i\sigma^{+-}\gamma_5}]}(x,\Blue{\bm{p}_T}) &=&
\frac{M}{P^+}\left(\lambda \Red{h_L(x,\Blue{\bm{p}_T^2})}
+\frac{\bm{p}_T\!\cdot\! \bm{S}_T}{M} \,\Red{h_T(x,\Blue{\bm{p}_T^2})}
\right)
\end{eqnarray}
\rule{0mm}{0mm}
\end{minipage}} \end{center} 

%% ---------   Slide 8 -------  spectator model I
%%%%%%%%%%%%%%%%%%%%%%%%%%%%%%%%%%%%%%%%%%%%%%%%%%%%%%%%%%%%%%%%%%%%%%%%
\section{spectator model}

%%%%%%%%%%%%%%%%%%%%%%%%%%%%%%%%%%%%%%%%%%%%%%%%%%
The purpose of our investigation is to obtain estimates for the non-leading
(i.e.~`twist 3') distribution functions which are experimentally 
poorly (or not at all) known at present. 

To this end we employ a rather simple spectator model with only a few
parameters. After fixing the parameters by phenomenological constraints we
check that the gross features of the experimentally well-known 
leading `twist' distribution functions $f_1(x)$ and $g_1(x)$ are
satisfactorily reproduced.

%%%%%%%%%%%%%%%%%%%%%%%%%%%%%%%%%%%%%%%%%%%%%%%%%%
\subsection{ingredients of the model}

The ingredients of the model are indicated below~\footnote{for more detailed
information about the model see also previous
publications~\cite{jmr97,mm91,tho94,nh95}}. 

\begin{itemize} 
\item
The basic idea of the spectator model is to assign a {\em definite
mass} to the {\em intermediate states} occurring in the quark-quark 
correlation functions \\[-4mm] 
\begin{center}
\includegraphics[width=4.5cm]{phi.eps}\\[1mm]
$(P-p)^2={M_R}^2$
\end{center}

\item
The {\em quantum numbers} of the intermediate state are determined
by the action of the quark field operator on the hadronic state $|P,S\rangle$,
i.e. they are the quantum numbers of a {\em diquark} system\\[-8mm]
\begin{center}\fbox{\begin{tabular}{rcc}
     & \ul{spin} & \ul{isospin} \tstrut\\
     scalar diquark & {0} & {0} \tstrut\\
axialvector diquark & {1} & {1}
\end{tabular}}
\end{center}

\item
The {\em matrix element} appearing in the RHS of (\ref{eq:Phi}) is
given by 
\begin{equation} 
\langle X_{s} | \psi_i(0) | P, S \rangle =
\left(i \over \pslash -m\right)_{ik} \;\Upsilon^{s}_{kl} 
\ U_l(P,S)
\end{equation} 
in the case of a scalar diquark, or by
\begin{equation} 
\langle X_{a}^{\lambda} | \psi_i(0) | P, S \rangle =
\epsilon_\mu^{*\lambda}
\left(i \over \pslash -m\right)_{ik} \;\Upsilon^{a\mu}_{kl} 
\ U_l(P,S)
\end{equation} 
for a vector diquark. The matrix elements consist of a nucleon-quark-diquark 
vertex $\Upsilon (N)$, the Dirac spinor
for the nucleon $U_l(P,S)$, a quark propagator for the untruncated quark line
and a polarization vector $\epsilon_\mu^{*(\lambda)}$
in the case of an axial vector diquark. 

\item
For the {\em nucleon-quark-diquark vertex} we assume the following
Dirac structures \footnote{a special case of the most general
form given in~\cite{tho94}.}:\\[2mm]  
\begin{minipage}{50mm} 
\includegraphics[width=3cm]{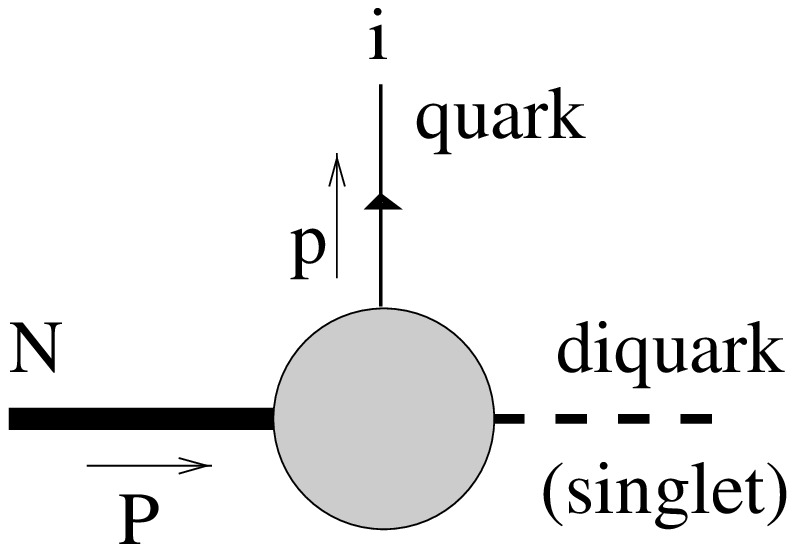}
\end{minipage} 
\begin{minipage}{110mm} 
\begin{equation}  \Upsilon^s (N) = \bm{1} \ g_s(p^2)\end{equation} 
\end{minipage}\\[4mm]
\begin{minipage}{50mm}  
\includegraphics[width=3cm]{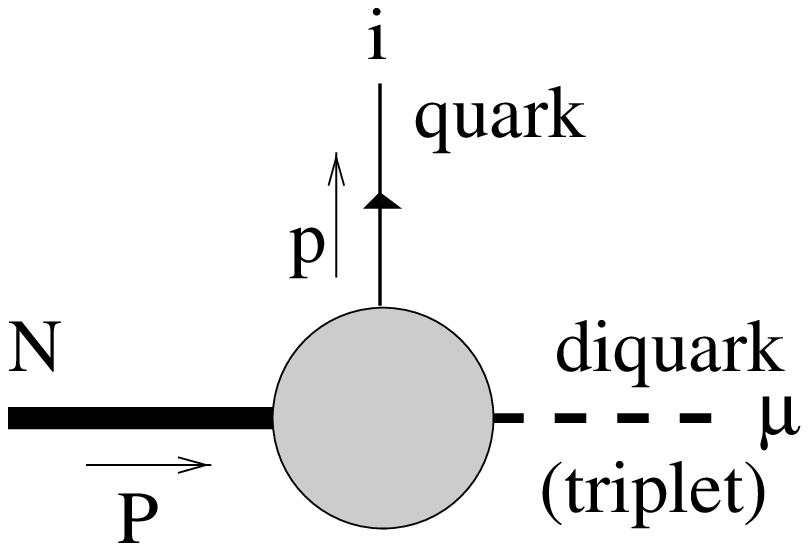}
\end{minipage} 
\begin{minipage}{110mm} 
\begin{equation} 
\Upsilon^{a \mu} (N) = \frac{g_a(p^2)}{\sqrt{3}}\,\gamma_5\,
                        \left(\gamma^\mu+\frac{P^\mu}{M}\right)
\end{equation} 
\end{minipage}

\item
The functions 
$g_{s/a} (p^2)$ are {\em form factors} that take into 
account the composite structure of the nucleon and the diquark 
spectator~\cite{tho94}. We use the same form factors for scalar and axial
vector diquark: 
\begin{equation} 
\label{eq:formfactors}
g_{s/a}(p^2) = N\,\frac{p^2 - m^2}{\left| p^2 - \Lambda^2 \right|^\alpha}.
\end{equation} 

\item
The {\em flavor} coupling of the proton wave function from a scalar diquark
($S_0$) and an (axial)vector diquark with isospin component $I_3=0$ or $1$ 
($A_0$ or $A_1$, respectively) 
\begin{equation} 
|p \rangle = {1 \over \sqrt{2}}\ \vert u \ S_0\rangle +
{1 \over \sqrt{6}} \vert u \ A_0\rangle
- {1\over \sqrt{3}} \ \vert d \ A_1\rangle,
\end{equation} 
leads to the flavor relations
\begin{eqnarray} 
f_1^u &=& \frac{3}{2}\,f_1^s + \frac{1}{2}\,f_1^a \\
f_1^d &=& f_1^a
\end{eqnarray} 
and similarly for the other functions. The coupling of the spin has already
been included in the vertices. 
\end{itemize} 
Putting all ingredients together analytic expression for the quark-quark
correlation functions are obtained
\begin{eqnarray} 
\Phi^R(p,P,S)={N^2\over 2(2\pi)^3}
{\delta(p^2-2P\cdot p+M^2-M_R^2)\over|p^2-\Lambda^2|^{2\alpha}}\nonumber\\ 
&&\qquad\qquad\times
(\pslash +m)(\Pslash+M)\left(1+a_R\gamma_5\Sslash\right)(\pslash +m)
\end{eqnarray} 
from which the distribution functions can be easily projected.
%%%%%%%%%%%%%%%%%%%%%%%%%%%%%%%%%%%%%%%%%%%%%%%%%%
\subsection{fixing the parameters}

The parameters of the model are fixed as follows: the power in the
denominator of the form factor, $\alpha=2$, is chosen to reproduce the 
Drell-Yan-West relation for large $x$. The mass difference $M_a-M_s$ is
motivated by the $N-\Delta$ mass difference (with group theoretical factors
properly accounted for) and the values for $\Lambda$ and $M_s$ reproduce the
experimental value for the axial charge, $g_A$.\\[-8mm]
\begin{center} \fbox{
\begin{tabular}{c|l}
parameter & fixed by: \tstrut \\ \hline\hline
{$\alpha=2$} & Drell-Yan-West relation:~$f_1^u(x)\sim (1-x)^3$\tstrut \\
{$M_a-M_s=200\;{\rm MeV}$} & $N-\Delta$ mass difference \tstrut \\
{$M_s=600\;{\rm MeV}; \Lambda=0.5\;{\rm GeV}$} & $g_A=1.26 $ \tstrut \\
{$N$} \tstrut & \parbox[t]{60mm}{number sum rules\\ 
                 $\int dx f_1^u(x)=2$ ; $\int dx f_1^d(x)=1$} 
\rule[-4mm]{0mm}{6mm}
\end{tabular}}
\end{center}

%%%%%%%%%%%%%%%%%%%%%%%%%%%%%%%%%%%%%%%%%%%%%%%%%%%%%%%%%%%%%%%%%%%%%%%%
\section{numerical results}

Having fixed the parameters of the model we can present our numerical
results. 

\begin{itemize}

%% ---------   Slide . -------
\item
\begin{minipage}[t]{50mm} 
The transverse momentum dependent distribution $f_1^u(x,\bm{p}_T)$
as obtained from the analytical expression for $\Phi$ by tracing with
$\gamma^+$ and integration over $dp^-$. The dependence on $\bm{p}_T$ is
driven by the choice of the form factors Eq.(\ref{eq:formfactors}) and 
momentum conservation.
\end{minipage} 
\hspace{4mm}
\begin{minipage}[t]{96mm} 
\vspace{-4mm}
\begin{center} 
\fbox{\includegraphics[width=9cm]{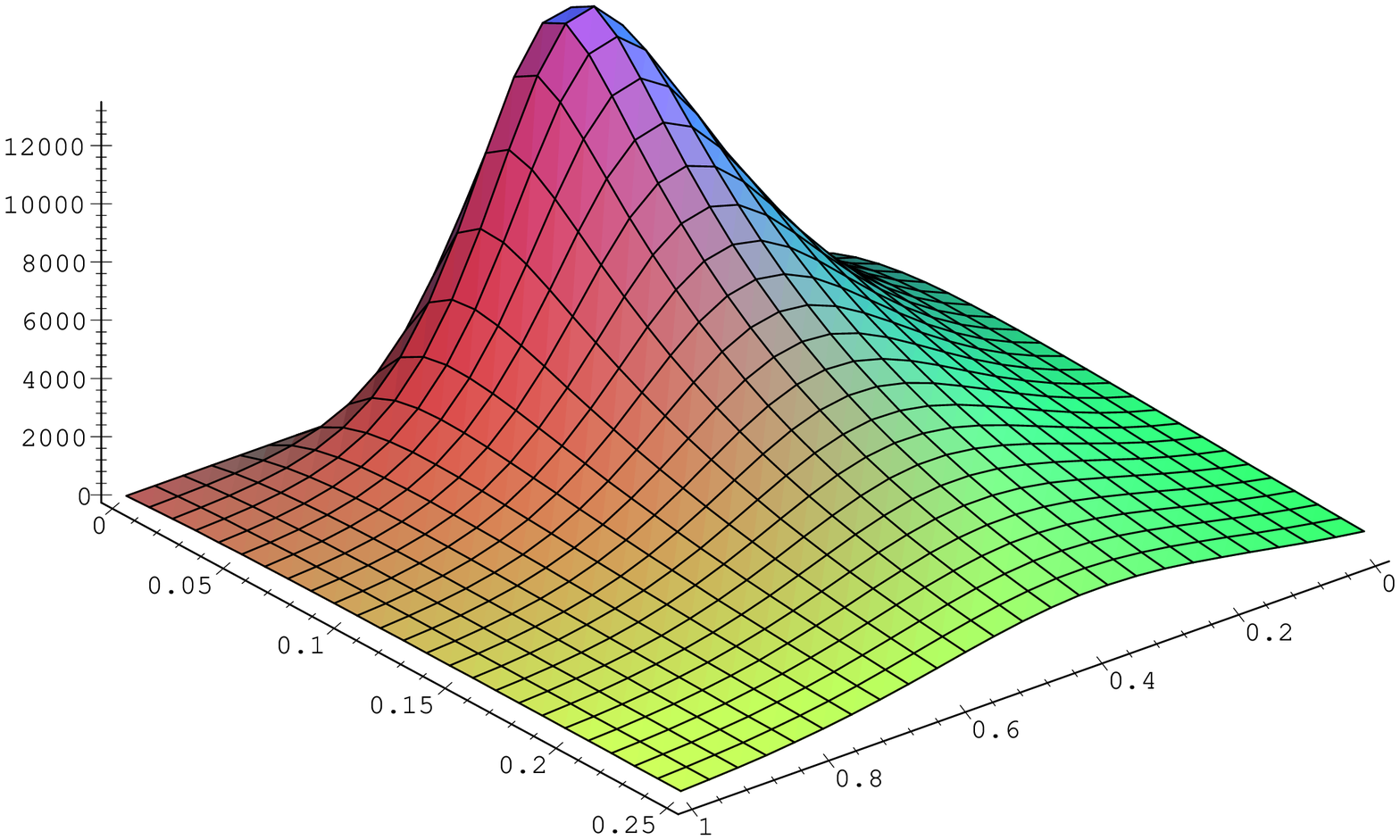}}
\end{center} 
\unitlength=1mm
\begin{picture}(0,0) 
\put(70,20){$x$}
\put(22,20){$\bm{p}_T^2$}
\end{picture}
\end{minipage}

%% ---------   Slide . -------

\item
Integrating $f_1(x,\bm{p}_T)$ over $d^2\bm{p}_T$ we obtain
the usual distribution $f_1(x)$. The values for the first moments
$\int_0^1 dx\,x\,f_1(x)$ are $0.690$ and $0.256$ for the $u$ and $d$
quark, respectively. We compare our result with the parametrization from
Gl\"uck, Reya and Vogt~\cite{grv95}  (at the low scale $\mu_{LO}^2=0.23\;{\rm
GeV}^2$)\footnote{Note that the model does 
not provide a scale dependence; but it is expected to describe physics at a low
``hadronic'' scale. Thus we compare to distributions found in the
literature at scales as low as available.}. 
\begin{center} \fbox{
\includegraphics[width=15cm]{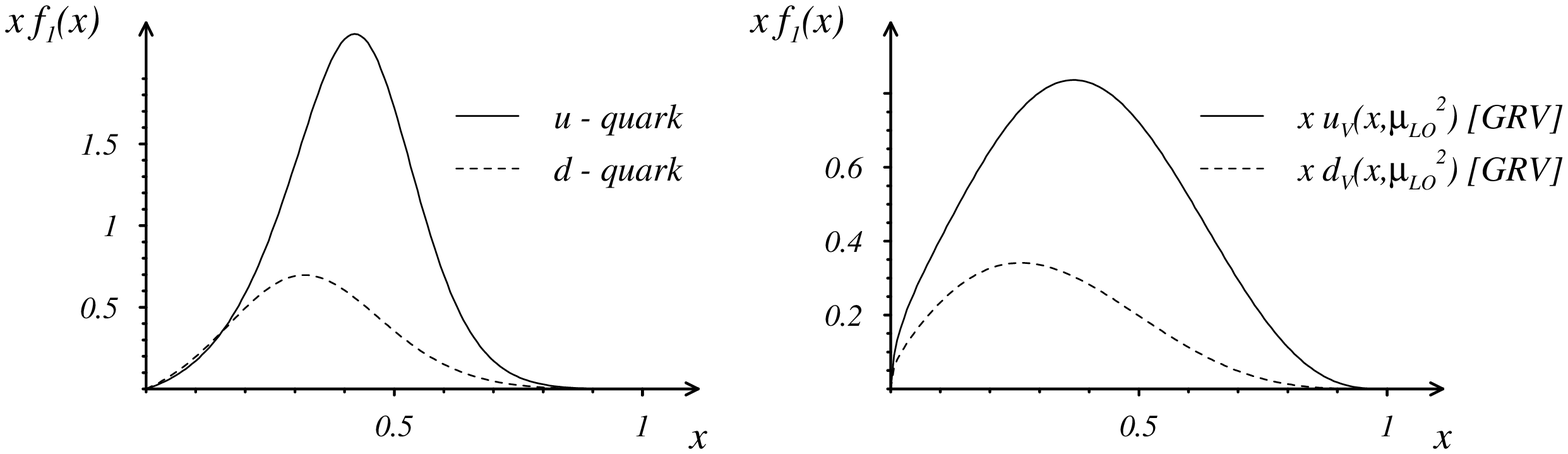}}
\end{center}
Note that the lowest moments, $\int f_1(x) dx$ are exactly the same 
(normalization condition), a fact not immediately apparent from the diagram,
since we plot the combination $x*f_1(x)$. The position of the maxima is in
fair agreement. Our distribution is narrower due to
the non-inclusion of gluons and anti-quarks 
(which -- if included  in our model -- would have a broadening effect). Thus,
we refrain from fine-tuning parameters to obtain a closer agreement with GRV
--- and are satisfied with agreement in the gross features.
%% ---------   Slide . -------
\item
The comparison of our result for $g_1(x)$ with parametrizations taken from the
literature~\cite{grsv96} reveals a similar agreement in the gross
features~\cite{jmr97}. 
\begin{center} \fbox{
\includegraphics[width=15cm]{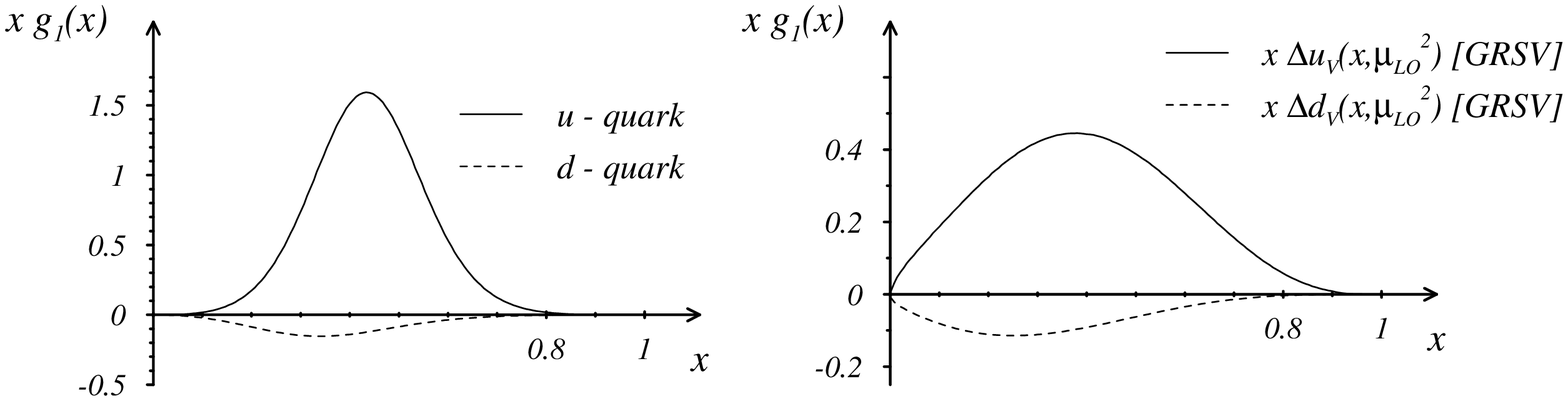}}
\end{center}
% ---------   Slide . -------

\item
Having gained some confidence in the model predictions -- as well as insight in
limitations in accuracy -- we can make predictions for the less well known 
or unknown functions.
\begin{center} 
\begin{minipage}[b]{85mm} 
\begin{center} 
\fbox{\includegraphics[width=7.9cm]{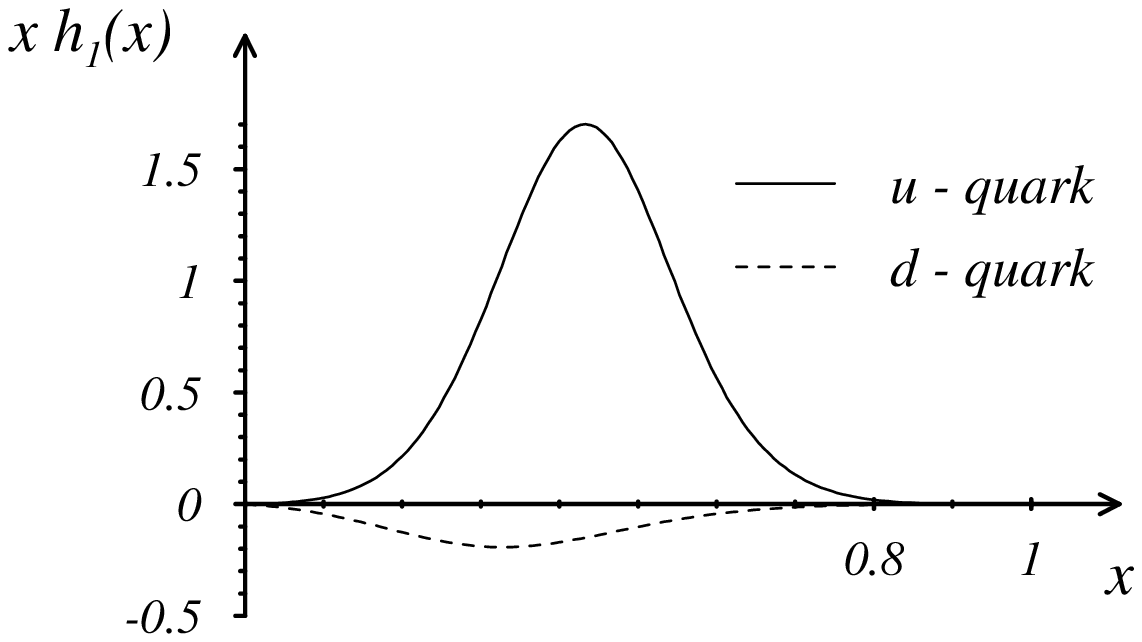}}
\end{center}
\end{minipage} 
\hspace{10mm}
\begin{minipage}[b]{55mm}
Yet experimentally completely undetermined is the transverse spin distribution
$h_1(x)$. Within our model we obtain an estimate for this function which is
numerically close to the result for the helicity distribution $g_1(x)$, but
not identical. 
\end{minipage} 
\end{center} 

%% ---------   Slide . -------

\item
We now turn to the `twist 3' distribution functions. We display 
$e(x)$ and the combination $g_2(x)=g_T(x)-g_1(x)$. The combination 
$h_2(x)=2(h_L(x)-h_1(x))$ is predicted in the
model to be just $2\times g_2(x)$.
\begin{center} 
\fbox{\includegraphics[width=15cm]{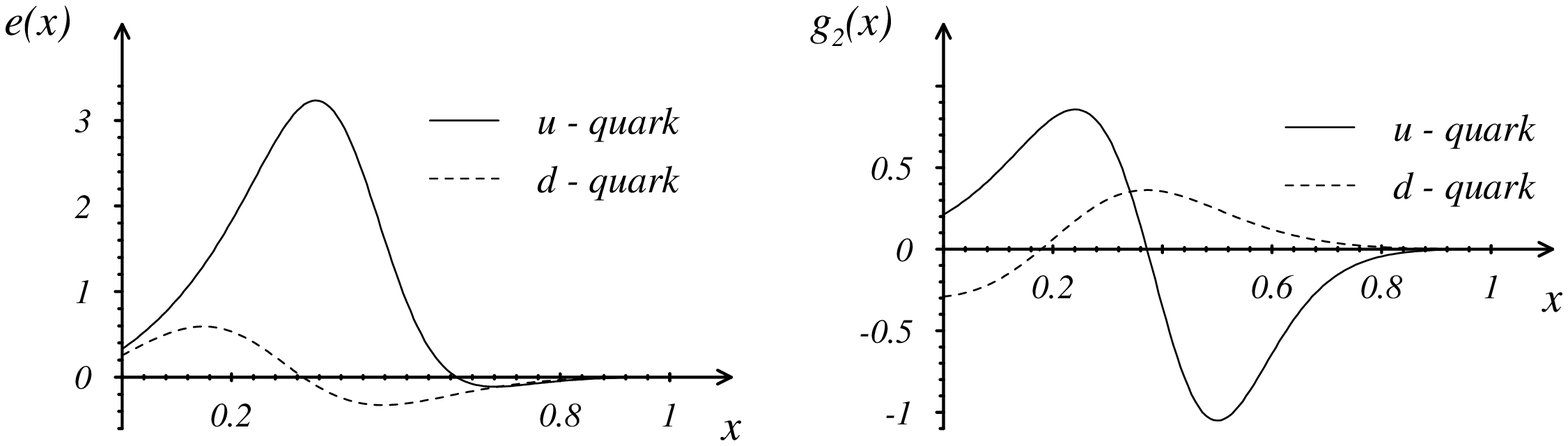}}
\end{center} 

%% ---------   Slide . -------

\item
From hermiticity and invariance under parity transformation relations between 
`leading twist' and `subleading twist' functions can be found
like~\cite{bkl84,tm96} 
\begin{eqnarray} 
g_T(x) &=& g_1(x) + \frac{d}{dx}\,g_{1T}^{(1)}(x)\\
&&\nn\\
h_L(x) &=& h_1(x) - \frac{d}{dx}\,h_{1L}^{\perp (1)}(x)
\end{eqnarray} 

\begin{minipage}{80mm} 
involving $\bm{p}_T$-moments of distribution functions
\[
g_{1T}^{(1)}(x)\,\equiv
\int d^2\bm{p}_T\,\left(\frac{\bm{p}_T^2}{2M^2}\right)
\,g_{1T}(x,\bm{p}_T)
\]
which, as well, are predicted by the model.\\[2mm]
The $\bm{p}_T$-moment
$h_{1L}^{\perp(1)}(x)$ is given by the relation 
$h_{1L}^{\perp(1)}(x)=-g_{1T}^{(1)}(x)$ within this model.\\[2mm]
 Note the 
non-vanishing values of $g_{1T}^{(1)}$ (and correspondingly of 
$h_{1L}^{\perp(1)}$) at $x=0$. 
\end{minipage} 
\begin{minipage}{80mm} 
\begin{center} 
\fbox{\includegraphics[width=7.6cm]{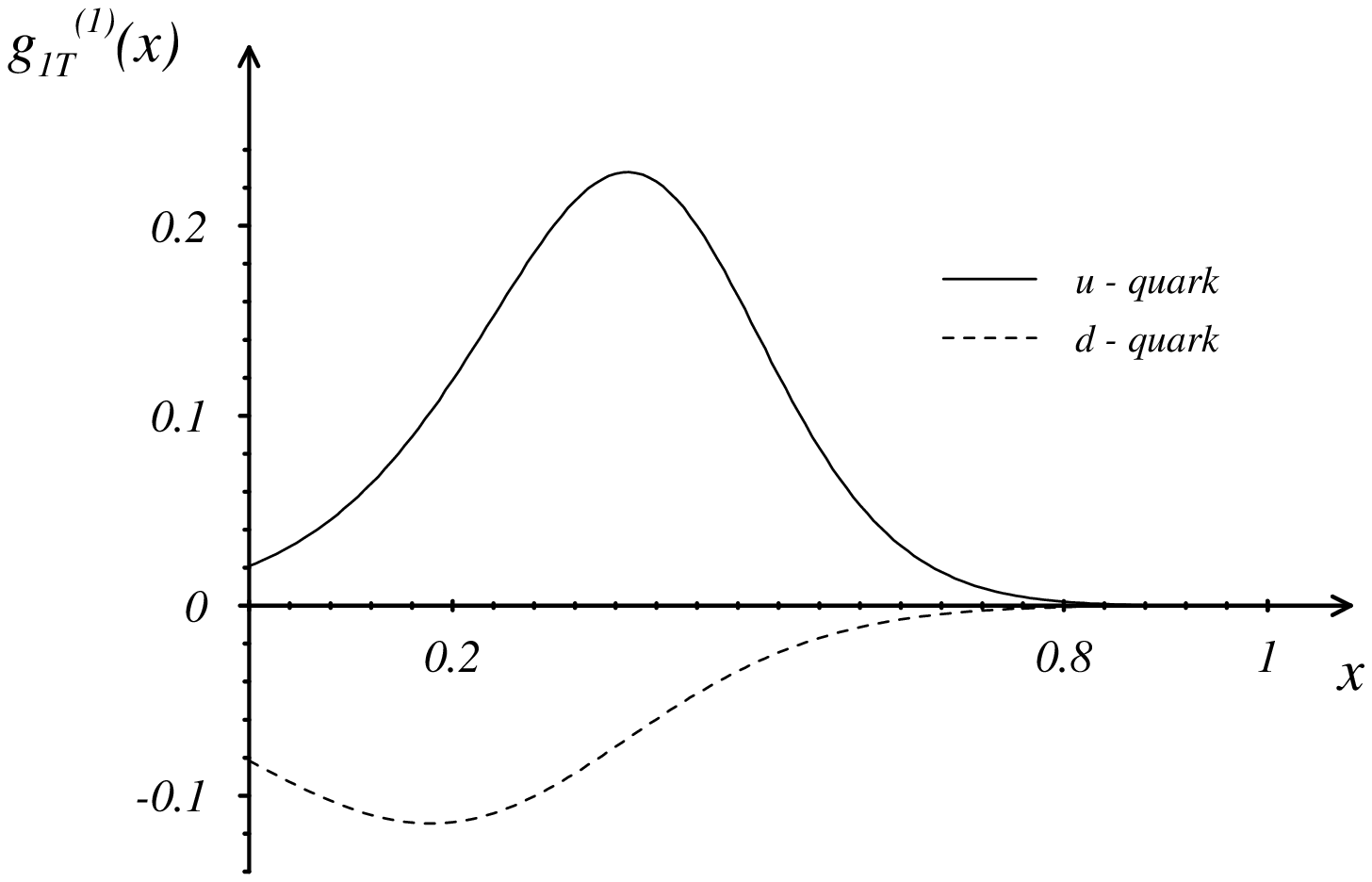}}
\end{center} 
\end{minipage}

\end{itemize}

%% ---------   Slide . -------

The positivity constraints 
$|g_1(x)|\leq f_1(x)$ and $|h_1(x)|\leq f_1(x)$
are trivially fulfilled in the model, as well as the Soffer~\cite{sof95}
inequality, $2|h_1(x)|\leq \left(f_1(x)+g_1(x)\right)$.\\[2mm]
On the other hand, the non-zero values of $g_{1T}^{(1)}(x=0)$ and
$h_{1L}^{\perp(1)}(x=0)$ indicate a small violation 
of the Burkhardt-Cottingham~\cite{bc70} sum rule, $\int_0^1 dx\,g_2(x)=0$,
and the Burkardt~\cite{bur92} sum rule, $\int_0^1 dx\,h_2(x)=0$. Those
violations turn out to be proportional to quark mass effects. Also, the
Efremov-Teryaev-Leader~\cite{etl97} sum rule, 
$\sum_{a\in V}e_a^2 \int_0^1 dx\,x\, \left(g_1^{a}(x)+2g_2^{a}(x)\right)=0$,  
is violated. Probably, those violations are artefacts of the model 
and might be used to constrain future refinements.

%%%%%%%%%%%%%%%%%%%%%%%%%%%%%%%%%%%%%%%%%%%%%%%%%%%%%%%%%%%%%%%%%%%%%%%
\section{conclusions}

\noindent
In the framework of a diquark spectator model we have obtained analytical
expressions for quark-quark correlation functions defined as hadronic matrix
elements of non-local operators. Distribution functions are given as
Dirac projections of the correlators. Numerical results for leading and, in
particular, sub-leading `twist' functions have been reported and discussed.

%%%%%%%%%%%%%%%%%%%%%%%%%%%%%%%%%%%%%%%%%%%%%%%%%%%%%%%%%%%%%%%%%%%%%%%
\section*{Acknowledgments}
This work is part of the scientific program of the
foundation for Fundamental Research on Matter (FOM),
the Dutch Organization for Scientific Research (NWO)
and the TMR program ERB FMRX-CT96-0008.


\begin{thebibliography}{99}
\negspace
\bibitem{jmr97}
R.~Jakob, P.J.~Mulders, J.~Rodrigues, NIKHEF 97-018, hep-ph/9704335.
\negspace
\bibitem{sop77}
D.E.~Soper, {\it Phys.~Rev.} {\bf D15} (1977) 1141; 
{\it Phys.~Rev.~Lett.} {\bf 43} (1979) 1847.
\negspace
\bibitem{cs82}
J.C.~Collins and D.E.~Soper, {\it Nucl.~Phys.} {\bf B194} (1982) 445. 
\negspace
\bibitem{jaf83}
R.L.~Jaffe, {\it Nucl.~Phys.} {\bf B229} (1983) 205.
\negspace
\bibitem{jj92}
R.L.~Jaffe, X.~Ji, {\it Nucl.~Phys.} {\bf B375} (1992) 527
\negspace
\bibitem{tm96}
R.D.~Tangerman, P.J.~Mulders, {\it Nucl.~Phys.} {\bf B461} (1996) 197
\negspace
\bibitem{mm91}
H.~Meyer and P.J.~Mulders, {\it Nucl.~Phys.} {\bf A528} (1991) 589.
\negspace
\bibitem{tho94}
W.~Melnitchouk, A.W.~Schreiber and A.W.~Thomas,
{\it Phys.~Rev.}~{\bf D49} (1994) 1183.
\negspace
\bibitem{nh95}
M.~Nzar and P.~Hoodbhoy, {\it Phys.~Rev.} {\bf D51} (1995) 32.
\negspace
\bibitem{grv95}
M.~Gl\"{u}ck, E.~Reya and A.~Vogt, {\it Z.~Phys.} {\bf C67} (1995) 433.
\negspace
\bibitem{grsv96}
M.~Gl\"{u}ck, {\it et al.}, {\it Phys.~Rev.} {\bf D53} (1996) 4775. 
\negspace
\bibitem{bkl84}
A.P.~Bukhvostov, E.A.~Kuraev, L.N.~Lipatov, {\it Sov.~Phys. JETP} {\bf 60}
(1984) 22.
\negspace
\bibitem{sof95}
J.~Soffer, {\it Phys.~Rev.~Lett.} {\bf 74} (1995) 1292.
\negspace
\bibitem{bc70}
H.~Burkhardt, W.N.~Cottingham, {\it Ann.~Phys.~(N.Y.)} {\bf 56} (1970) 453.
\negspace
\bibitem{bur92}
M.~Burkardt, {\sf Proceedings of 10th International Symposium on High Energy
Physics, Nagoya, Japan, 1992} (Universal Academy Press, Tokyo, 1993)
\negspace
\bibitem{etl97} 
A.V.~Efremov, O.V.~Teryaev, E.~Leader, {\it Phys.~Rev.} {\bf D55} (1997) 4307.
\end{thebibliography}
\end{document}